\documentclass[final,pra,aps,twocolumn,showpacs]{revtex4}
\usepackage{amsfonts}
\usepackage{amsmath}
\usepackage{amssymb}
\usepackage[english]{babel}
\input{epsfx}
\usepackage[dvipdf]{graphicx}
\usepackage{capt-of}
\usepackage{indentfirst}
\newcommand{\ket}[1]{\ensuremath{\left|#1\right\rangle}}
\newcommand{\ud}{\mathrm{d}}

\begin{document}
\title{Multiphoton resonance in a three~-~level system with nearly degenerate excited states}
\author{M.~Berent}
\email{mberent@amu.edu.pl}
\author{R.~Parzy\'nski}
\affiliation{Faculty of Physics, Adam Mickiewicz University, Umultowska 85, 61-614 Pozna\'n, Poland}
\date{\today}

\begin{abstract}
An analytic study is presented of the efficient multiphoton excitation and strong harmonic generation in three-level systems specified by a pair of nearly degenerate, strongly dipole-coupled excited states. Such systems are physically formed by the three lowest states in, e.g., the hydrogen atom or evenly charged homonuclear diatomic molecular ions under reasonably chosen laser intensities. As a detailed analytic result, we found that the laser pulse of photon energy $2{.}05\text{ eV}$, duration $0{.}23\text{ ps}$ and intensity $5\cdot 10^{13}\,\frac{\text{W}}{\text{cm}^2}$ is able to produce complete inversion of the initial population in the hydrogen atom through the 5-photon excitation. At the same photon energy, the pulse of duration $0{.}41\text{ ps}$ and intensity $3{.}44\cdot 10^{14}\,\frac{\text{W}}{\text{cm}^2}$ was found to produce the same effect in the molecular ion but through the 9-photon excitation. We show that the accompanying scattering of light has very rich spectrum differing substantially from that of the two-level system.
\end{abstract}


\maketitle
\section{Introduction}
\label{sec:intr}
Typically, a number of optical photons are necessary to cover the energy gap between the ground state and the first excited state in most atoms and molecules. Multiphoton transition is thus the only way of populating the excited states in these materials when exposed to laser light. In a nonperturbative analysis, Duvall \textit{et.al.} \cite{duv} have, however, proved that in a two-level system multiphoton excitation is inefficient because of small multiphoton coupling strengths requiring high laser fields producing, in turn, large ac Stark shifts and prompting competitive processes like multiphoton ionization. As shown by Gibson \cite{gib2002,gib2003}, all these disadvantages of the two-level system are cancelled in a specific class of three-level systems, where the multiphoton coupling strengths are greatly enhanced, the ac Stark shifts are greatly reduced and the ionization is minimized. The specific three-level systems of interest include a pair of nearly-degenerate strongly-coupled different-parity excited states separated by many photons from the ground state. A physical realization of such systems are, e.g., all evenly charged homonuclear diatomic molecular ions, like $N_2^{4+}$ for example, as reminded by Gibson. The recent numerical calculations by Gibson for a 1D model of this ion have shown that its ground state is distanced by as much as over $0{.}66\text{ a.u.}$ (over $18\text{ eV}$) from the excited pair of states at the internuclear separation of the order of $3{.}5\text{ a.u.}$. At this internuclear separation, the splitting of the excited states amounts to $0{.}0167\text{ a.u.}$ ($0{.}454\text{ eV}$) and the dipole matrix element between them is 6 times greater than that between the ground state and the lower excited state. By numerical calculations Gibson found that, at laser intensities of the order of $10^{15}\,\frac{\text{W}}{\text{cm}^2}$, nearly complete inversion of the population could be obtained in the 11-photon and 12-photon excitation processes. Also, an important finding was that no other discrete ionic states, besides the three lowest ones, were populated during the excitation and practically no ionization occurred at intensities not exceeding substantially $10^{15}\,\frac{\text{W}}{\text{cm}^2}$. All this means that, up to the mentioned laser intensities, the ion behaves like a perfect three-level system. Thanks to the efficient multiphoton excitation, this ion was predicted by Gibson to be a source of strong generation of harmonics of the incident laser light. At the above intensities, the harmonic generation by the ion differs in its mechanism from the generation by atoms because in the latter case the atomic continuum is strongly engaged through the three-step mechanism including ionization, acceleration of the ionized electron and, finally, its recombination \cite{lew}. In the case of atoms, e.g., the hydrogen atom with its three lowest states $1S$, $2P$ and $2S$, a substantial decrease in the laser intensity by almost two orders of magnitude is required to avoid the ionization effects and to approximate the atom by an effective three-level system.\\
\indent In the present paper we give an approximate analytic solution to the problem of the above specific three-level systems strongly driven by a laser field. Our analytic solution will be shown to confirm the numerically-based predictions of Gibson concerning efficient multiphoton excitation and strong harmonic generation. The chain of dipole couplings in the system under the present study forms a $\Gamma$-type configuration (see Fig.\ref{fig:sys}). This configuration differs from the $\Lambda$-type and ladder-type configurations we have previously studied \cite{parz} under the opposite assumption of weak multiphoton excitation and by a different method based on a set of two differential equations for appropriately defined ratios of the level population amplitudes. At weak excitations, this set was made decoupled and one equation was transformed into a quadratically nonlinear Riccati-type equation. Thus, the present paper considerably extends our previous treatment by including a different system, a different approach and higher laser intensities.

\section{Theory}
\label{sec:intro}
\subsection{The model}

\begin{figure}[!ht]
\centering 
\includegraphics[width=0.9\columnwidth]{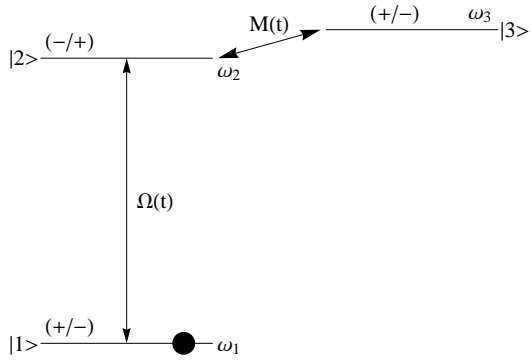}
\caption{The three-level $\Gamma$-type system with nearly degenerate upper states in a laser field of frequency $\omega_{32}\ll\omega_0\ll\omega_{21}$. $\Omega(t)$ and $M(t)$ are instantaneous Rabi frequencies for the 1-2 and 2-3 dipole couplings, respectively. The dot points to the initial population.}\label{fig:sys}
\end{figure}

The system of interest, shown in Fig.\ref{fig:sys} , includes three states \ket{j} ($j = 1,2,3$) of eigenfrequencies $\omega_j$ and defined parities (either ''+'' or ''-''). The parities of states \ket{1} and \ket{3} are assumed to be the same and opposite to that of state \ket{2}. This system, with its initial population in \ket{1}, is driven by a light of linear polarization along the $z$-axis, carrier frequency $\omega_0$, electric field amplitude $E_0$ and switching on (off) function $0\leq f(t) \leq 1$. The interaction is taken in the electric-dipole approximation $V(t) = -e z E(t)$, where $E(t) = E_0 f(t) \cos(\omega_0 t)$. Instead of $V_{ij}(t)$, we introduce the instantaneous Rabi frequencies for the $1-2$ and $2-3$ couplings, namely $\Omega (t) = -\frac{V_{12}(t)}{\hbar} = \Omega_R f(t) \cos(\omega_0 t)$ and $M (t) = -\frac{V_{23}(t)}{\hbar} = M_R f(t) \cos(\omega_0 t)$, where $\Omega_R = \frac{\mu_{12} E_0}{\hbar}$ and $M_R = \frac{\mu_{23} E_0}{\hbar}$ are the standard  time-independent Rabi frequencies  while $\mu_{jk} = \langle j|e z|k\rangle$ is the dipole matrix element for the $j\rightarrow k$ transition. 
Throughout this paper the duration of the light pulse is taken to be much shorter than the excited state lifetimes, so the function of the system under interaction is represented by $\Psi(t) = \sum_{j=1}^{3} b_j(t) |j\rangle$. From the Schr\"odinger equation for $\Psi(t)$ one obtains the differential equations for the time-dependent population amplitudes $b_j(t)$:
\begin{eqnarray}
i \dot{b}_1 &=& \omega_1 b_1 - \Omega(t) b_2\,,\nonumber\\
i \dot{b}_2 &=& \omega_2 b_2 - \Omega(t) b_1 - M(t) b_3\,,\nonumber\\
i \dot{b}_3 &=& \omega_3 b_3 - M(t) b_2\,.\label{eq:bamp}
\end{eqnarray}
\indent We perform a series of transformations on this set. Four of them are those suggested by Gibson~\cite{gib2003} for the system with degenerate upper states ($\omega_2=\omega_3$). In turn, the first transformation is
\begin{equation}
c_j = b_j\, e^{i \omega_2 t}\,,
\end{equation}
the second
\begin{eqnarray}
c_1 &=& c_1\,,\nonumber\\
c_{\pm} &=& \frac{1}{\sqrt{2}}(c_2\pm c_3)\,,
\end{eqnarray}
the third
\begin{eqnarray}
d_1 &=& c_1\, e^{-i \omega_{21} t}\,,\nonumber\\
d_{\pm} &=& c_{\pm}\, e^{\mp i \phi(t)}\,,
\end{eqnarray}
and the fourth
\begin{eqnarray}
d_1 &=& d_1\,,\nonumber\\
d_{2,3} &=& \frac{1}{\sqrt{2}}(d_{+}\pm d_{-})\,,
\end{eqnarray}
where $\omega_{\alpha \beta} = \omega_{\alpha}-\omega_{\beta}$, $\phi(t) = \int_{t_0 = 0}^{t}M(t^{'}) \ud t^{'}$ and $t_0$ is the turn-on time of the pulse. To these transformations we add one more transformation
\begin{eqnarray}
x &=& d_1\,,\nonumber\\
y &=& d_2\, \exp\left[i \omega_{32}\int_{0}^{t}\sin^2(\phi(t^{'}))\ud t^{'}\right]\,,\nonumber\\
z &=& d_3\, \exp\left[i \omega_{32}\int_{0}^{t}\cos^2(\phi(t^{'}))\ud t^{'}\right]\,.
\end{eqnarray}
By these transformations the initial set of equations for $b_j$ is replaced by a new set for $x$, $y$ and $z$:
\begin{eqnarray}
i \dot{x} &=& F(t) y + G(t) z\,,\nonumber\\
i \dot{y} &=& F^{*}(t) x + H(t) z\,,\nonumber\\
i \dot{z} &=& G^{*}(t) x + H^{*}(t) y\,,\label{eq:set1}
\end{eqnarray}
where
\begin{eqnarray}
F(t) &=& - \exp\left[-\frac{i}{2}(\omega_{21}+\omega_{31})t\right] q(t) \Omega(t)\! \cos(\phi(t))\label{eq:wspf},\\
G(t) &=& -i \exp\!\!\left[\!-\frac{i}{2}(\omega_{21}\!+\!\omega_{31})t\!\right]\!\! q^{-1}(t) \!\Omega(t) \!\sin(\phi(t))\label{eq:wspg},\\
H(t) &=& -i \frac{\omega_{32}}{2} q^{-2}(t) \sin(2 \phi(t))\label{eq:wsph}
\end{eqnarray}
with
\begin{equation}
q(t) = \exp\left[\frac{i}{2} \omega_{32}\int_{0}^{t} \cos(2 \phi(t^{'})) \ud t^{'}\right]\label{eq:q}\,.
\end{equation}
Obviously, set (\ref{eq:set1}) for $x$, $y$ and $z$ is as exact as set~(\ref{eq:bamp}) for $b_j$. The advantage of the set for $x$, $y$ and $z$ is that it makes it possible to identify multiphoton resonances in the system. These resonances are rooted in the coupling terms $F(t)$, $G(t)$ and $H(t)$ being nonlinear in $M(t)$.\\
\indent Having found $x$, $y$ and $z$ we return to the amplitudes $b_j$ by the inverse transformations
\begin{eqnarray}
b_1 &=& e^{-i \omega_1 t} x\,,\label{eq:tr1}\\
b_2 &=& - e^{-i \omega_1 t} \Omega^{-1}(t) \left(F(t) y + G(t) z\right)\,,\label{eq:tr2}\\
b_3 &=& - e^{-i\omega_1 t} \Omega^{-1}(t) \!\left(q^{-2}(t)F(t) z\! +\! q^2(t) G(t) y \right).\label{eq:tr3}
\end{eqnarray}
From the initial conditions for $b_j(t)$ ($b_1(0) = 1$, $b_2(0) = b_3(0) =0$), one obtains the following initial conditions for  $x$, $y$ and $z$, namely $x(0) = 1$, $y(0) = z(0) = 0$. We will use the above amplitudes to calculate the population dynamics, $|b_j(t)|^2$, and the coherent part of the correspondence-principle spectrum of scattered light. The total spectrum, $S(\omega)$, is defined by \cite{lappas}
\begin{equation}
S(\omega)/\omega^4 = \sum_{j=1}^{3}\left|\int_{0}^{t_p} \ud t\,e^{i \omega t} \langle\Psi_j(t)|e z|\Psi_1(t)\rangle \right|^2\,,\label{eq:spec}
\end{equation}
where $t_p$ is the duration of the incoming pulse and $\Psi_j(t)$ stands for the Schr\"odinger wave function satisfying the condition $\Psi_j(0) = |j\rangle$. The diagonal term in Eq.(\ref{eq:spec}), $j = 1$, defines the  coherent part of the spectrum, $S_c(\omega)$, and the rest is the incoherent part, $S_{inc}(\omega)$. The coherent part results from the average dipole moment $d(t) = \langle\Psi_1(t)|e z|\Psi_1(t)\rangle = 2 \mu_{12} Re\left(b^{*}_{1}(t)b_2(t)\right)+ 2 \mu_{23} Re\left(b_{2}^{*}(t)b_3(t) \right)$, while the incoherent part from the dipole fluctuations around the average value. In all our calculations we will take many-cycle pulse, i.e., much longer than the optical period $T = \frac{2\pi}{\omega_0}$, meaning the pulse shape function $f(t)$ to be slowly varying in time when compared with $\cos(\omega_0 t)$. This assumption will allow us to remove the slow function $f(t)$ from the integrand when multiplied by a fast function. Thanks to that, $\Omega(t)\cos(\phi(t))$ and $\Omega(t)\sin(\phi(t))$, being the major parts of $F(t)$ and G(t), respectively, can be expanded in harmonic functions as
\begin{eqnarray}
\Omega(t)\cos(\phi(t)) &=& 2 \omega_0 \frac{\Omega_R}{M_R}\sum_{k=0}^{\infty} (2 k+1)J_{2 k+1}\left(\frac{M_R}{\omega_0}f(t)\right)\nonumber\\
&\cdot& \cos\left[(2 k+1)\omega_0 t\right]\,,\\
\Omega(t)\sin(\phi(t)) &=& 2 \omega_0 \frac{\Omega_R}{M_R}\sum_{k=1}^{\infty} (2 k)J_{2 k}\left(\frac{M_R}{\omega_0}f(t)\right)\nonumber\\
&\cdot& \sin\left(2 k\omega_0 t\right)\,.
\end{eqnarray}
The above expansions result from the known relations for the Bessel functions $J_k(x)$ \cite{ryzhik}:
\begin{eqnarray}
\cos(\rho \sin\alpha) &=& J_0(\rho) + 2 \sum_{k=1}^{\infty} J_{2 k}(\rho) \cos(2 k \alpha)\,,\label{eq:fb1}\\
\sin(\rho \sin\alpha) &=& 2 \sum_{k=0}^{\infty} J_{2 k+1}(\rho) \sin((2k+1)\alpha)\,,\label{eq:fb2}
\end{eqnarray}
\begin{equation}
J_{k-1}(\rho) + J_{k+1}(\rho) = \frac{2k}{\rho}J_k(\rho)\,.
\end{equation}


\subsection{Exact degeneracy of the excited states ($\omega_{32} = 0$)}
\label{sec:3}

\indent In the case of exact degeneration of the upper levels, $q(t) = 1$ and $H(t) = 0$ and then set (\ref{eq:set1}) reduces to
\begin{eqnarray}
i \dot{x} &=& F(t) y + G(t) z\,,\nonumber\\
i \dot{y} &=& F^{*}(t) x\,,\nonumber\\
i \dot{z} &=& G^{*}(t) x\,,\label{eq:set2}
\end{eqnarray}
where now
\begin{eqnarray}
F(t) &=& - \omega_0 \frac{\Omega_R}{M_R} \sum_{n=1,3,5,\dots} n J_n\left(\frac{M_R}{\omega_0} f(t) \right)\nonumber\\
&\cdot&\left(e^{i(n\omega_0 - \omega_{21})t} + e^{-i(n\omega_0 + \omega_{21})t} \right)\label{eq:F}\,,\\
G(t) &=& - \omega_0 \frac{\Omega_R}{M_R} \sum_{p=2,4,6,\dots} p J_p\left(\frac{M_R}{\omega_0} f(t) \right)\nonumber\\
&\cdot&\left(e^{i(p\omega_0 - \omega_{31})t} - e^{-i(p\omega_0 + \omega_{31})t} \right)\label{eq:G}\,.
\end{eqnarray}
When $M_R\rightarrow 0$, we have $F(t) = -\Omega(t) e^{-i\omega_{21}t}$ and $G(t) = 0$ and then equations (\ref{eq:set2}) are made the standard equations for the two~-~level system under one-photon resonance.\\
\indent As seen, $F(t)$ determines the odd~-~photon excitation of \ket{2}, while $G(t)$ - the even-photon excitation of \ket{3}. Under the odd-photon resonance ($\omega_{21} = N \omega_0+\delta_N$, $|\delta_N|\ll \omega_0$, $N = 1,3,5,\cdots$), $F(t)$ splits into the slow part $F_N$ and the rapidly oscillating part $f_N$, while $G(t) = G_N$ always remains a rapidly varying function. Precisely:
\begin{equation}
F_N = - N \omega_0 \frac{\Omega_R}{M_R} J_N\left(\frac{M_R}{\omega_0}f(t)\right) e^{-i\delta_N t} = - a_N e^{-i\delta_N t}\,,\label{eq:rabio}
\end{equation}

\begin{widetext}
\begin{eqnarray}
f_N &=& -\omega_0 \frac{\Omega_R}{M_R} e^{-i\delta_N t } \left\{\sum_{\begin{subarray}{c}
n = 1,3,5,\cdots\\
n\neq N
\end{subarray}}n J_n\left(\frac{M_R}{\omega_0}f(t)\right) e^{i(n-N)\omega_0 t}+\sum_{n = 1,3,5,\cdots}n J_n\left(\frac{M_R}{\omega_0}f(t)\right) e^{-i(n+N)\omega_0 t} \right\}\,,\label{eq:fastf}
\end{eqnarray}
\end{widetext}

\begin{eqnarray}
G_N &=&  -\omega_0 \frac{\Omega_R}{M_R} e^{-i\delta_N t } \sum_{p=2,4,6,\cdots} p J_p\left(\frac{M_R}{\omega_0}f(t)\right)\nonumber\\
&\cdot& \left(e^{i(p-N)\omega_0 t} - e^{-i(p+N)\omega_0 t} \right)\,.\label{eq:fastg}
\end{eqnarray}
In the other case of even~-~photon resonance ($\omega_{31} = P\omega_0 +\delta_P$, $|\delta_P| \ll \omega_0$, $P = 2,4,6,\cdots$), the functions $F(t)$ and $G(t)$ change their roles. Now, $G(t)$ is composed of its slow part $G_P$ and the fast part $g_P$, while $F(t) = F_P$ remains a rapidly varying function:
\begin{equation}
G_P = - P \omega_0 \frac{\Omega_R}{M_R} J_P\left(\frac{M_R}{\omega_0}f(t)\right) e^{-i\delta_P t} = - a_P e^{-i\delta_P t}\,,\label{eq:rabie}
\end{equation}

\begin{widetext}
\begin{eqnarray}
g_P &=& -\omega_0 \frac{\Omega_R}{M_R} e^{-i\delta_P t } \Bigg(\sum_{\begin{subarray}{c}
p = 2,4,6,\cdots\\
p\neq P
\end{subarray}} p J_p\left(\frac{M_R}{\omega_0}f(t)\right) e^{i(p-P)\omega_0 t}-\sum_{p = 2,4,6,\cdots}p J_p\left(\frac{M_R}{\omega_0}f(t)\right) e^{-i(p+P)\omega_0 t} \Bigg)\,,
\end{eqnarray}
\end{widetext}

\begin{eqnarray}
F_P &=&  -\omega_0 \frac{\Omega_R}{M_R} e^{-i\delta_P t } \sum_{n = 1,3,5,\cdots} n J_n\left(\frac{M_R}{\omega_0}f(t)\right)\nonumber\\
&\cdot& \left(e^{i(n-P)\omega_0 t} + e^{-i(n+P)\omega_0 t} \right)\label{eq:29}\,.
\end{eqnarray}
The slow functions $F_N$ and $G_P$ are seen to have the same structure and become constant for exact multiphoton resonance ($\delta_N = \delta_P = 0$) caused by a square pulse ($f(t) = 1$). These constants are the resonant multiphoton Rabi frequencies identified by Gibson earlier (Eq. (14) in \cite{gib2003}). On the other hand, the fast functions oscillate with either even multiplicities ($f_N$, $g_P$) or odd multiplicities ($G_N$, $F_P$) of the carrier frequency $\omega_0$ of the incident light. Under a given multiphoton resonance, equations (\ref{eq:set2}) are transformed into
\begin{eqnarray}
i \dot{x}_N &=& (F_N+f_N) y_N + G_N z_N\,,\nonumber\\
i \dot{y}_N &=& (F_N^{*}+f_N^{*}) x_N\,,\nonumber\\
i \dot{z}_N &=& G_N^{*} x_N\label{eq:set3}
\end{eqnarray}
for the odd-photon excitation ($N = 1,3,5,\cdots$), and
\begin{eqnarray}
i \dot{x}_P &=& F_P y_P + (G_P+g_P) z_P\,,\nonumber\\
i \dot{y}_P &=& F_P^{*} x_P\,,\nonumber\\
i \dot{z}_P &=& (G_P^{*}+g_P^{*}) x_P\label{eq:set4}
\end{eqnarray}
for the even-photon excitation ($P = 2,4,6,\cdots$). Thus far, no approximation (besides the assumption of slow pulse envelope) has been made, when manipulating the equations.\\
\indent In the lowest-order (rough) approximation, we are tempted to neglect all rapidly oscillating terms in equations (\ref{eq:set3}) and (\ref{eq:set4}) in complete analogy to the standard rotating wave approximation (RWA) known from the one-photon resonance \cite{allen,scully}. In such an approach, equations (\ref{eq:set3}) are reduced to $i \dot{x}_N = - a_N e^{-i \delta_N t} y_N$, $i \dot{y}_N = -a_N e^{i \delta_N t} x_N$, $i \dot{z}_N = 0$, while equations (\ref{eq:set4}) to $i \dot{x}_P = - a_P e^{-i \delta_P t} z_P$, $i \dot{z}_P = - a_P e^{i\delta_P t}x_P$, $i \dot{y}_P = 0$. In either case, the first two equations are structurally the same as those for the two-level system under RWA, but now with the multiphoton couplings $a_K$ ($K = N,P$) defined by equations (\ref{eq:rabio}) and (\ref{eq:rabie}). Obviously, in the limit $M_R\rightarrow 0$, only $a_1$ takes a non~-~zero value $\left(a_K\rightarrow \frac{\Omega_R f(t)}{2}\delta_{K,1}\,\text{where }\delta_{a,b}\text{ is the Kronecker symbol} \right)$. Choosing, for example, a square pulse ($f(t) = 1$) we find
\begin{eqnarray}
x_N(t) &=& \Bigg(x_N(0)\cos(A_N t)+\frac{i}{A_N}\Big(\frac{\delta_N}{2}x_N(0)\nonumber\\&+&a_N y_N(0) \Big)\sin(A_N t) \Bigg) e^{-i\frac{\delta_N}{2} t}\,,\nonumber\\
y_N(t) &=& \Bigg(y_N(0)\cos(A_N t)-\frac{i}{A_N}\Big(\frac{\delta_N}{2}y_N(0)\nonumber\\&-&a_N x_N(0) \Big)\sin(A_N t) \Bigg) e^{i\frac{\delta_N}{2} t}\,,\nonumber\\
z_N(t) &=& z_N(0)\,,\label{eq:sol1}
\end{eqnarray}
where $A_N = \sqrt{a_N^2+\left(\frac{\delta_N}{2}\right)^2}$ has the meaning of an off-resonant multiphoton (odd~-~photon) Rabi frequency.It results from the derivation that the solutions (\ref{eq:sol1}) have to change slowly on the scale of the field period, so the restriction of their applicability is $A_N\ll \omega_0$. For the even~-~photon resonance, $x_P$ is obtained from $x_N$ by changing $N\rightarrow P$ and $y\rightarrow z$ simultaneously; $z_P$ is obtained from $y_N$ in the same way. Obviously, $y_P(t) = y_P(0)$ in the latter case. For the initial conditions $b_1(0) = 1$, $b_2(0) = b_3(0) = 0$, equivalent to $x(0) = 1$, $y(0) = z(0) = 0$, we combine the solutions (\ref{eq:sol1}) with the transformation relations (\ref{eq:tr1})~--~(\ref{eq:tr3}) and obtain the time-dependent population amplitudes $b_j$. In the case of odd-photon resonance ($N = 1,3,5,\cdots$), these amplitudes are
\begin{eqnarray}
b_1^{(N)}(t) &=& \!\!\!\left[\cos(A_N t)\!+\!i\frac{\delta_N}{2 A_N}\sin(A_N t) \right]\!e^{i(N\omega_0+\frac{\delta_N}{2}-\omega_2)t},\nonumber\\
b_2^{(N)}(t) &=& i \frac{a_N}{A_N} \sin(A_N t)\cos(\phi(t)) e^{i(\frac{\delta_N}{2}-\omega_2 )t}\,,\nonumber\\
b_3^{(N)}(t) &=& - \frac{a_N}{A_N} \sin(A_N t)\sin(\phi(t)) e^{i(\frac{\delta_N}{2}-\omega_2 )t}\,.\label{eq:rwa}
\end{eqnarray}
For the even-photon resonance ($P = 2,4,6,\cdots$), one obtains $b_1^{(P)}$ from $b_1^{(N)}$,  $b_2^{(P)}$ from $b_3^{(N)}$ and  $b_3^{(P)}$ from $b_2^{(N)}$ by changing $N\rightarrow P$ in equations (\ref{eq:rwa}). At any time $t$, the above amplitudes satisfy the condition of normalization of the total population probability, $|b_1(t)|^2+|b_2(t)|^2+|b_3(t)|^2 = 1$. With these amplitudes we can analyze not only the population dynamics, $|b_j(t)|^2$, but the coherent spectrum of scattered light as well, $S(\omega)/\omega^4 = |\int_0^{t_p} d t e^{i\omega t} d(t)|^2$, where the average dipole moment is now reduced to $d(t) = \langle \Psi_1(t)|e z |\Psi_1(t)\rangle = 2\mu_{12} Re\left(b_1^{*}b_2 \right)$ because $Re\left(b_2^{*}b_3\right) = 0$ as results from (\ref{eq:rwa}). For $N$-photon excitation ($N = 1,3,5,\cdots$), we obtain from equations (\ref{eq:rwa}) and the Fourier-Bessel expansions (\ref{eq:fb1}), (\ref{eq:fb2}) the following induced dipole:
\begin{eqnarray}
d^{(N)}(t) &=& \mu_{12}\frac{a_N}{2 A_N}\sum_{k=0}^{\infty}\alpha_k J_{2 k}\left(\frac{M_R}{\omega_0}\right)\nonumber\\
&\cdot&\Big\{\frac{\delta_N}{A_N}\left[\cos(N-2 k)\omega_0 t +\cos(N+2 k)\omega_0 t\right]\nonumber\\
&+&\left(1-\frac{\delta_N}{2 A_N}\right)\left[\cos\left((N-2k)\omega_0 - 2 A_N\right)t\right.\nonumber\\
&+& \left.\cos\left((N+2k)\omega_0-2A_N\right)t \right]\nonumber\\
&-&\left(1+\frac{\delta_N}{2 A_N}\right)\left[\cos\left((N-2k)\omega_0 + 2 A_N\right)t\right.\nonumber\\
&+& \left.\cos\left((N+2k)\omega_0+2A_N\right)t \right] \Big\}\label{eq:dip}\,,
\end{eqnarray}
where $\alpha_0 = \frac{1}{2}$ and $\alpha_{k>0} = 1$, by definition. In the case of even-photon excitation ($P = 2,4,6,\cdots$), $d^{(P)}(t)$ is obtained from the above $d^{(N)}(t)$ in the following steps: first  we put $1$ in place of all $\alpha_k$, then replace $N\rightarrow P$ and $2 k\rightarrow 2 k+1$ and, finally, change the sign $+\rightarrow -$ at the second cosine in each bracket. These average dipoles suggest that odd-order harmonics will be revealed in the coherent spectrum $S_c(\omega)$ with each harmonic having a triplet structure, in general. At exact multiphoton resonance ($\delta_K=0$, where $K = N,P$), the triplets convert into doublets with the field-dependent separation $4 a_K$ between the doublet components. In the limit $M_R\rightarrow 0$, we recover from Eq. (\ref{eq:dip}) the known coherent spectrum, i.e. the doublet with separation $2 \Omega_R$ around the perfectly resonant incoming frequency $\omega_0 = \omega_{21}$.\\
\indent An improved solution to equations (\ref{eq:set3}) is found by applying the approach of Avetissian \textit{et al.} \cite{avet2002,avet2008}. According to this approach,  $x$, $y$ and $z$ are separated into their slow and rapid parts, e.g., $x(t) = \bar{x}(t)+\beta_x(t)$, where $\bar{x}(t)$ is the time average of $x(t)$ while $\beta_x(t)$ is a rapidly oscillating function on the scale of the incident field  period $T = \frac{2\pi}{\omega_0}$. Then, we substitute the two-part expressions for  $x$, $y$ and $z$ into (\ref{eq:set3}), separate the slow and rapid oscillations and neglect small terms $F_N \beta_y \ll \dot{\beta}_x$ and $F^{*}_N \beta_x \ll \dot{\beta}_y$ in equations for the rapid parts. The rapid parts are found to satisfy the set
\begin{eqnarray}
i\dot{\beta}_x &=& \bar{y}_Nf_N(t)+\bar{z}_N G_N(t)\,,\nonumber\\
i\dot{\beta}_y &=& \bar{x}_N f_N^{*}(t)\,,\nonumber\\
i\dot{\beta}_z &=& \bar{x}_N G_{N}^{*}(t)\,.\label{eq:rap1}
\end{eqnarray}
Having in mind that $\bar{x}_N$, $\bar{y}_N$ and $\bar{z}_N$ are slow functions with respect to $f_N$ and $G_N$, this set is easily integrated. When the result of this integration is substituted to the equations for the slow parts and then time-averaging of the right-hand sides is made, one obtains
\begin{eqnarray}
i\,\dot{\bar{x}}_N &=& \left(\Delta_{f_N}+\Delta_{G_N}\right)\bar{x}_N - a_N e^{-i\delta_N t}\bar{y}_N\,,\nonumber\\
i\,\dot{\bar{y}}_N &=& -\Delta_{f_{\!N}}\bar{y}_N - a_N e^{i\delta_N t}\bar{x}_N\,,\nonumber\\
i\,\dot{\bar{z}}_N &=& -\Delta_{G_{\!N}}\bar{z}_N\,,\label{eq:set5}
\end{eqnarray}
where
\begin{eqnarray}
\Delta_{f_N} &=& -i\overline{f_N(t)\int f_N^{*}(t^{'}) \ud t^{'}}\nonumber\\ &\stackrel{|\delta_N|\ll\omega_0}{=}& 2 N \omega_0 \left(\frac{\Omega_R}{M_R} \right)^2\nonumber\\
&\cdot&\sum_{\begin{subarray}{c}
n = 1,3,5,\cdots\\
n\neq N
\end{subarray}} \frac{n^2}{n^2-N^2}J_{n}^{2}\left(\frac{M_R}{\omega_0}f(t) \right)\label{eq:ourdf}\,,
\end{eqnarray}
\begin{eqnarray}
\Delta_{G_N} &=& -i\overline{G_N(t)\int G_N^{*}(t^{'}) \ud t^{'}}\nonumber\\ &\stackrel{|\delta_N|\ll\omega_0}{=}& 2 N \omega_0 \left(\frac{\Omega_R}{M_R} \right)^2\nonumber\\
&\cdot&\sum_{\begin{subarray}{c}
p = 2,4,6,\cdots
\end{subarray}} \frac{p^2}{p^2-N^2}J_{p}^{2}\left(\frac{M_R}{\omega_0}f(t) \right)\label{eq:ourdG}
\end{eqnarray}
have the meaning of the dynamic Stark shifts. By the substitutions $\bar{x}_N = u \exp\left[-i\left(\Delta_{f_N}+\Delta_{G_N} \right)t \right]$ and $\bar{y}_N = v \exp\left[i\Delta_{f_N}t \right]$, the first two equations in (\ref{eq:set5}) are reduced to $i\,\dot{u} = -a_N\exp\left[-i\bar{\delta}_N t \right]v$, $i\,\dot{v} = -a_N\exp\left[i\bar{\delta}_N t \right]u$, where $\bar{\delta}_N = \delta_N-\left(2\Delta_{f_N}+\Delta_{G_N} \right)$ is the Stark-modified $N$~-~photon detuning. The equations for $u$, $v$ have the same structure as the previously discussed set $i\,\dot{x}_N = -a_N\exp\left[-i\delta_N t \right]y_N$, $i\,\dot{y}_N = -a_N\exp\left[-i\delta_N t \right]x_N$. Thus,  $u$ and $v$ are obtained from the first two equations (\ref{eq:sol1}), by replacing $x_N\rightarrow u$, $y_N\rightarrow v$ and also $\delta_N\rightarrow \bar{\delta}_N$, $A_N\rightarrow \bar{A}_N = \sqrt{a_N^2+\left(\frac{\bar{\delta}_N}{2}\right)^2}$. As a result, the population dynamics, $\left|b_j^{(N)}\right|^2$, still results from equations (\ref{eq:rwa}) after the replacements $\delta_N\rightarrow \bar{\delta}_N$, $A_N\rightarrow \bar{A}_N$. The same replacements must be done in equation (\ref{eq:dip}) for the average dipole. With the above Stark~-~type corrections, the applicability range of the solutions is now $\bar{A}_N\ll\omega_0$. Some corrections to both the population dynamics and the average dipole are expected to come from the rapidly oscillating parts $\beta_x$, $\beta_y$ and $\beta_z$ of $x$, $y$ and $z$. These rapid parts are to be calculated from equations (\ref{eq:rap1}) after the slow parts $\bar{x}_N$, $\bar{y}_N$ and $\bar{z}_N$ have already been found. Obviously, what has been said above holds for the even~-~photon excitation as well.
\subsection{Near degeneracy of the excited states $\left(\left|\frac{\omega_{32}}{\omega_0} \right|\ll 1\right)$}\label{sec:4}
The theory of Section \ref{sec:3} can straightforwardly be adapted to the case of nearly degenerate upper states, i.e., those satisfying the condition $\left|\frac{\omega_{32}}{\omega_0} \right|\ll 1$. After using equation (\ref{eq:fb1}), $q(t)$ of equation (\ref{eq:q}) can be reduced to the form $q(t)=\exp\left(\frac{i}{2}\widetilde{\omega}_{32} t\right)$, where $\widetilde{\omega}_{32} =\omega_{32}J_0\left(\frac{2 M_R}{\omega_0}f(t)\right) = \omega_{32}-2\Delta(t)$ with $\Delta(t)=\frac{\omega_{32}}{2}\left(1-J_0\left(\frac{2 M_R}{\omega_0}f(t) \right) \right)$. As a result, $F(t)$ is still given by equation (\ref{eq:F}) and $G(t)$ by equation (\ref{eq:G}) with the replacements $\omega_{21}\rightarrow\widetilde{\omega}_{21} = \omega_{21}+\Delta(t)$ and $\omega_{31}\rightarrow\widetilde{\omega}_{31} = \omega_{31}-\Delta(t)$, respectively. Obviously, $H(t)$ of equation (\ref{eq:wsph}) differs from zero now but is a rapidly oscillating function due to (\ref{eq:fb2}) and our approximation for $q(t)$. In the case of multiphoton resonances , $\delta_N\rightarrow \widetilde{\delta}_N = \widetilde{\omega}_{21}-N \omega_0$ and $\delta_P\rightarrow \widetilde{\delta}_P = \widetilde{\omega}_{31}-P \omega_0$ in equations (\ref{eq:rabio})~-~(\ref{eq:fastg}) and (\ref{eq:rabie})~-~(\ref{eq:29}), respectively. Also, the term $H(t) z_N$ ($H^{*}(t)y_N$) must be added to the right-hand  side of the second (third) equation of the set (\ref{eq:set3}). Similar extension must be applied to the set (\ref{eq:set4}) but with the change $N\rightarrow P$ now.\\
\indent These changes only slightly modify the solutions of Section \ref{sec:3}. For example, making the rough RWA, consisting in neglecting the rapidly varying terms $f_N$, $G_N$ and $H_N$ in the extended set (\ref{eq:set3}), one arrives (for $f(t)=1$) at equations (\ref{eq:sol1}), (\ref{eq:rwa}) and (\ref{eq:dip}) with the change $\delta_N\rightarrow \widetilde{\delta}_N$. On the other hand, when applying the improved method of Avetissian \textit{et al.}, we again approach the set (\ref{eq:rap1}) in which $i\dot{\beta}_y$ has the extra term $\bar{z}_N H(t)$ on its right~-~hand side, while $i\dot{\beta}_z$ the extra term $\bar{y}_N H^{*}(t)$. Also,  we recover the set (\ref{eq:set5}) with $\delta_N\rightarrow\widetilde{\delta}_N$ and with some coefficients appropriately redefined. Precisely, $a_N\rightarrow a_N+\alpha_N$ in the first equation,  $a_N\rightarrow a_N+\alpha_N$ and $\Delta_{f_N}\rightarrow \Delta_{f_N}-\Delta_H$ in the second equation, and $\Delta_{G_N}\rightarrow \Delta_{G_N}+\Delta_H$ in the third equation, where
\begin{eqnarray}
\alpha_N &=&-i\overline{G_N(t)\int H^{*}(t^{'})\ud t^{'}}\nonumber\\
&\stackrel{|\omega_{32}|\ll\omega_0}{=}& \omega_{32}\frac{\Omega_R}{M_R}
\sum_{p=2,4,\cdots} J_p\left(\frac{M_R}{\omega_0}f(t)\right)\nonumber\\
&\cdot& J_{p-N}\left(\frac{2 M_R}{\omega_0}f(t)\right)\frac{p}{p-N}
\end{eqnarray}
is a correction to the coupling parameter $a_N$, while
\begin{eqnarray}
\Delta_H &\stackrel{|\omega_{32}|\ll\omega_0}{=}& \overline{-i H(t)\int H^{*}(t^{'})\ud t^{'}}\nonumber\\
&=& -\frac{1}{2}\left(\frac{\omega_{32}}{\omega_0}\right)^2\omega_{32}J_0\left(\frac{2 M_R}{\omega_0}f(t)\right)\nonumber\\
&\cdot&\sum_{n=1,3,\cdots}\frac{J_n^2\left(\frac{2 M_R}{\omega_0}f(t)\right)}{n^2}
\end{eqnarray}
is an additional Stark shift.
\section{Applications}
\begin{figure}[!ht]
\centering 
\includegraphics[width=0.9\columnwidth]{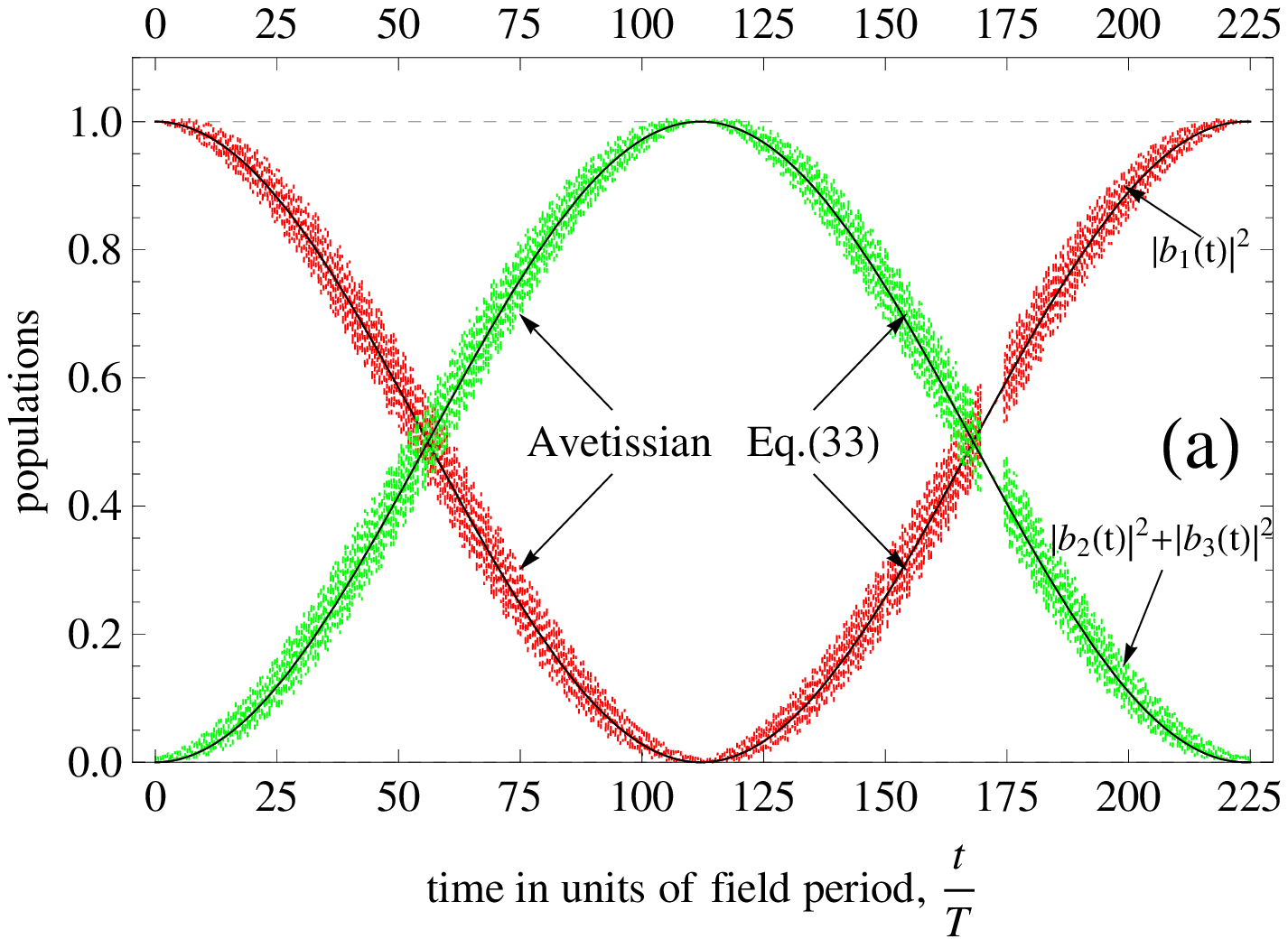}
\includegraphics[width=0.9\columnwidth]{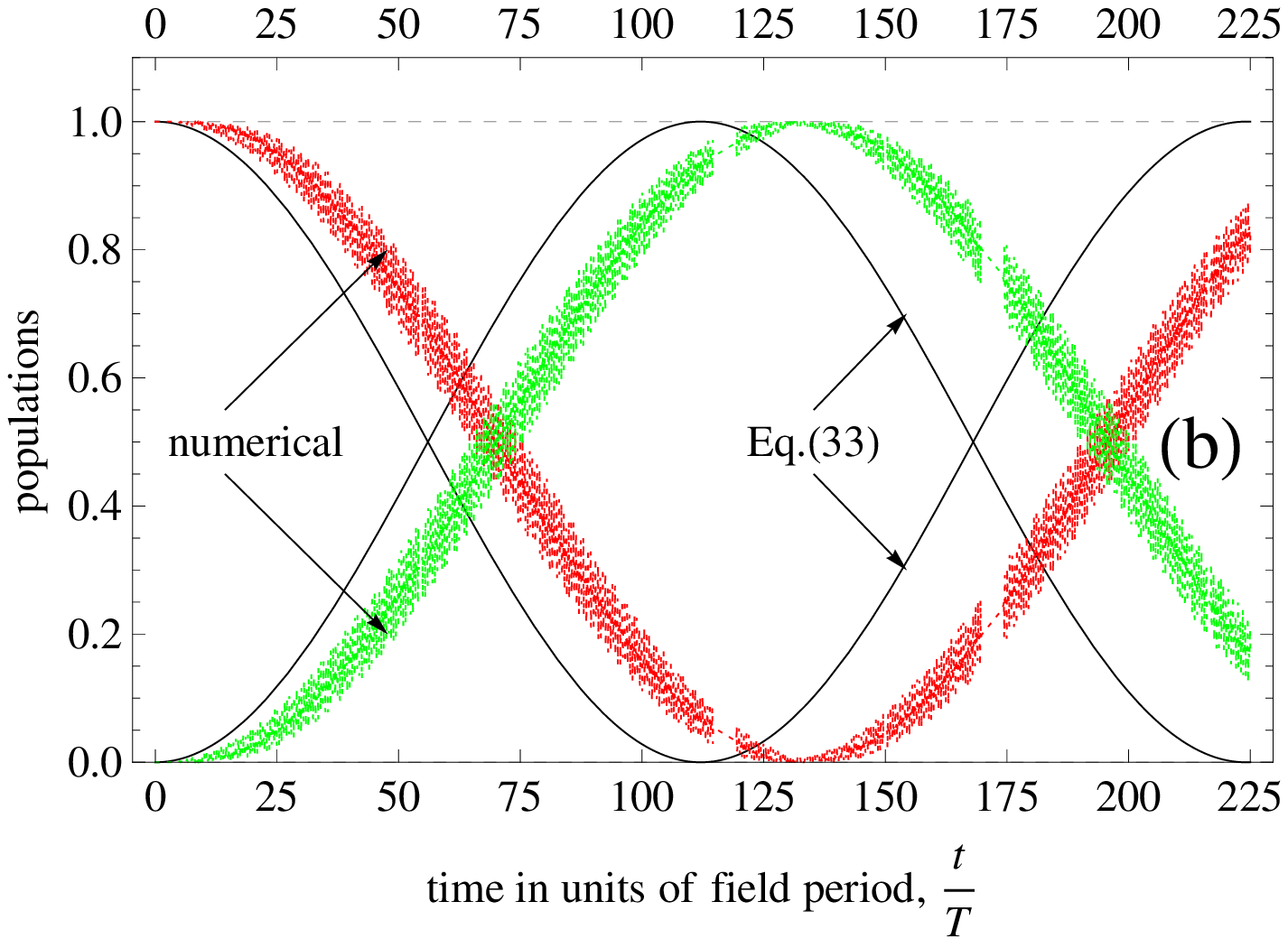}
\includegraphics[width=0.9\columnwidth]{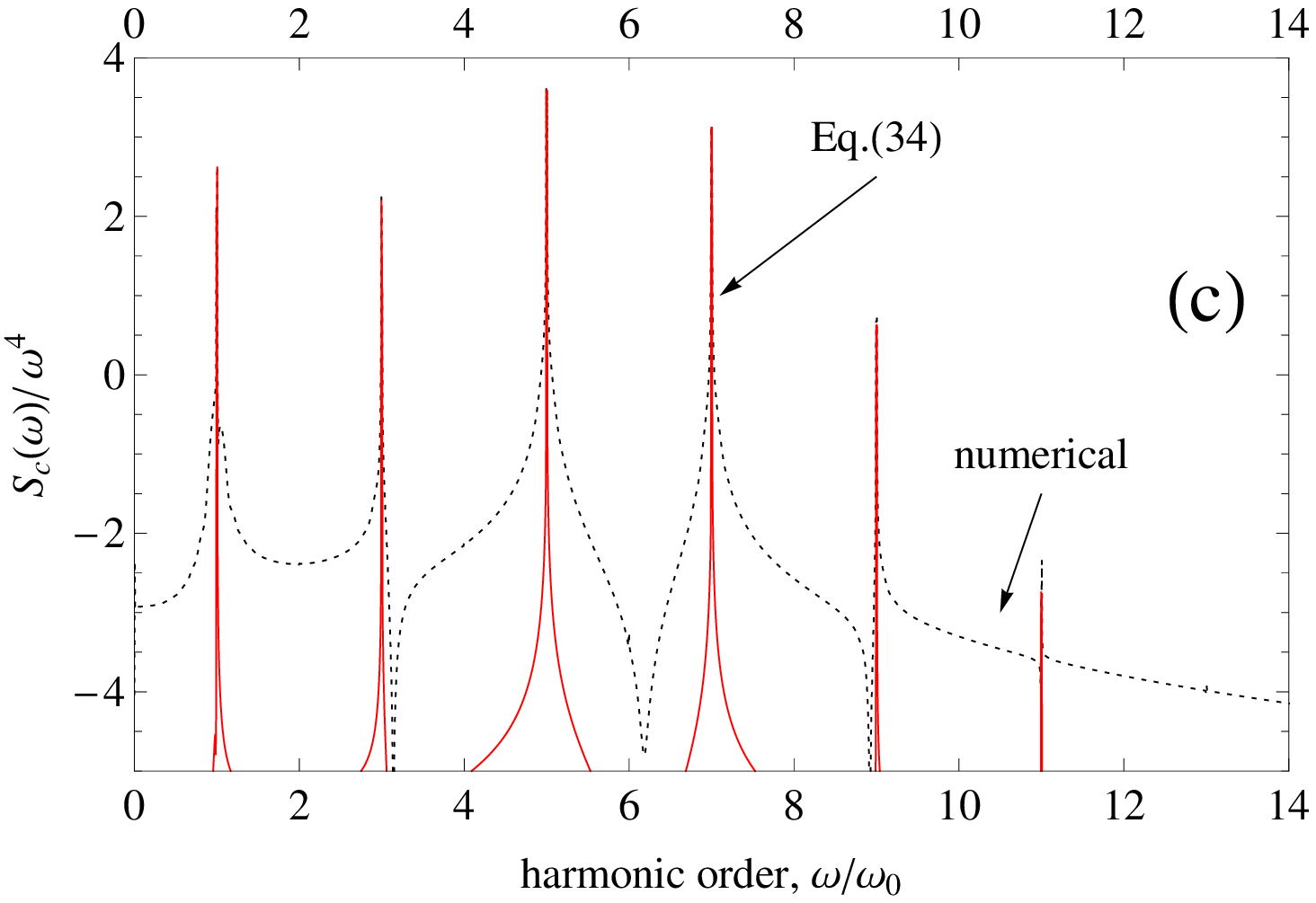}
\caption{(Color online) Five-photon ($N=5$) $\text{1S}\rightarrow\text{2P}$ resonance in the hydrogen atom produced by the laser field of strength $E_0 = 0{.}0377\text{ a.u.}$ and frequency $\omega_0 = 0{.}0754\text{ a.u.}$ (a) and (b), temporal evolutions of level populations. (c) coherent spectrum of scattered light . Bold solid curves correspond to approximate analytic solutions (either Eq.(\ref{eq:rwa}) or Eq.(\ref{eq:dip})). The small rapid oscillations superimposed on the bold curve in (a) come from the approach of Avetissian \textit{et.al.} while those in (b) come from numerical integration of Eq.(\ref{eq:bamp}) for the field with 10-cycle turn-on time and then of constant amplitude. The turn-on function was $\sin^2(\omega_0 t/40)$ with the restriction $\omega_0 t/40\leq\frac{\pi}{2}$. In (c), the numerical spectrum was obtained for the field with the same turn-on function as in (b).}\label{fig:2}
\end{figure}
\begin{figure}[!ht]
\centering 
\includegraphics[width=0.9\columnwidth]{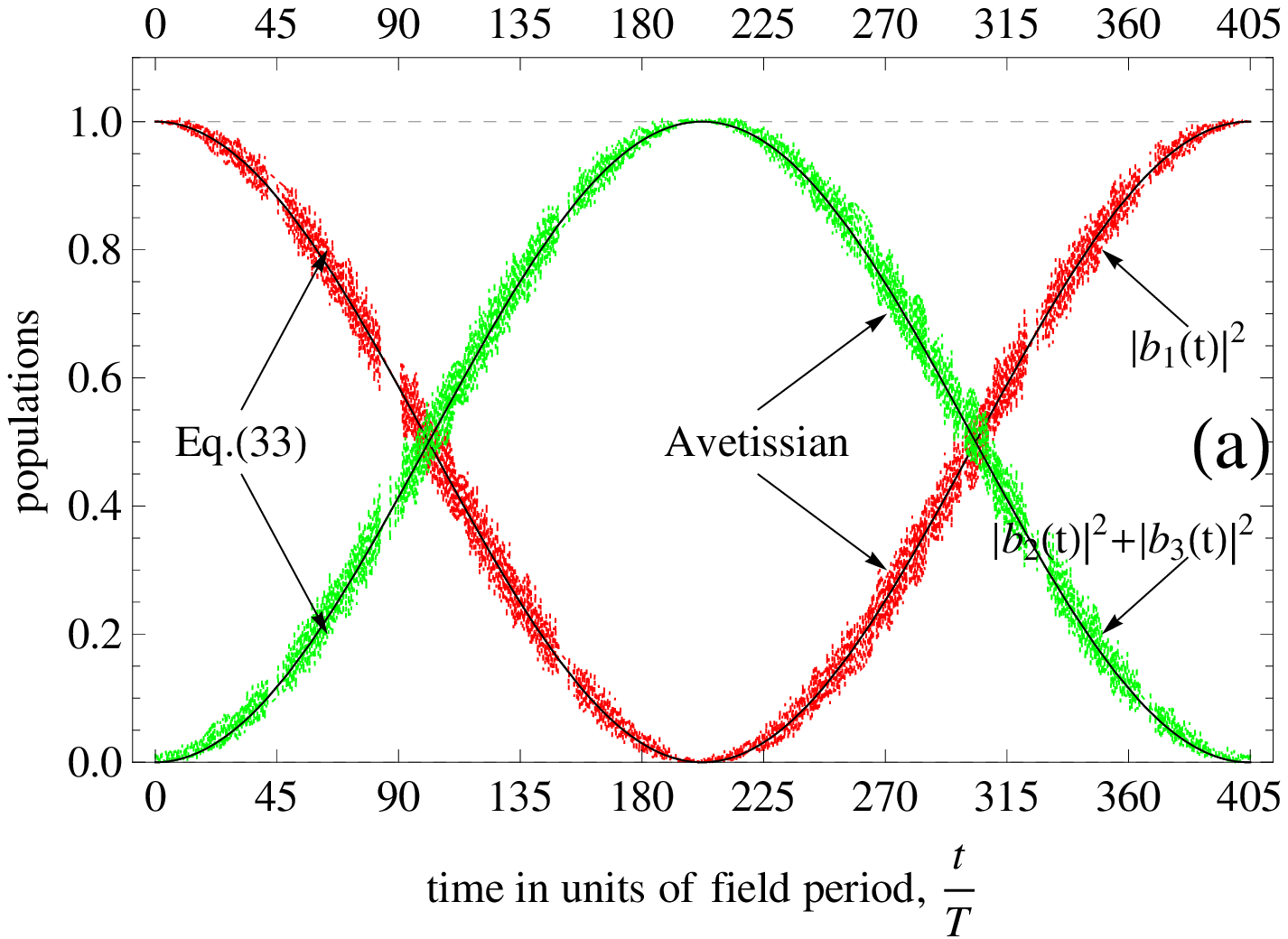}
\includegraphics[width=0.9\columnwidth]{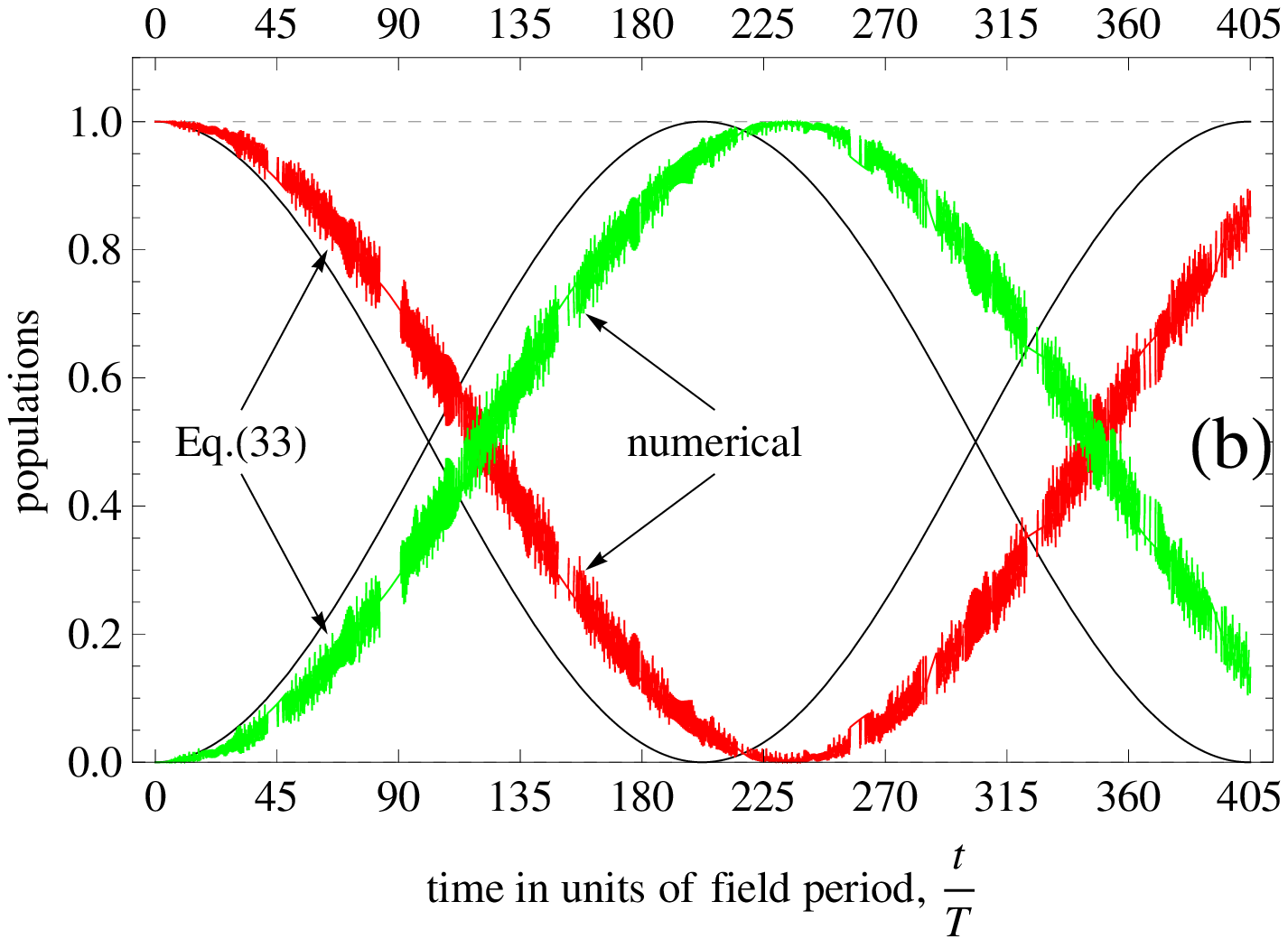}
\includegraphics[width=0.9\columnwidth]{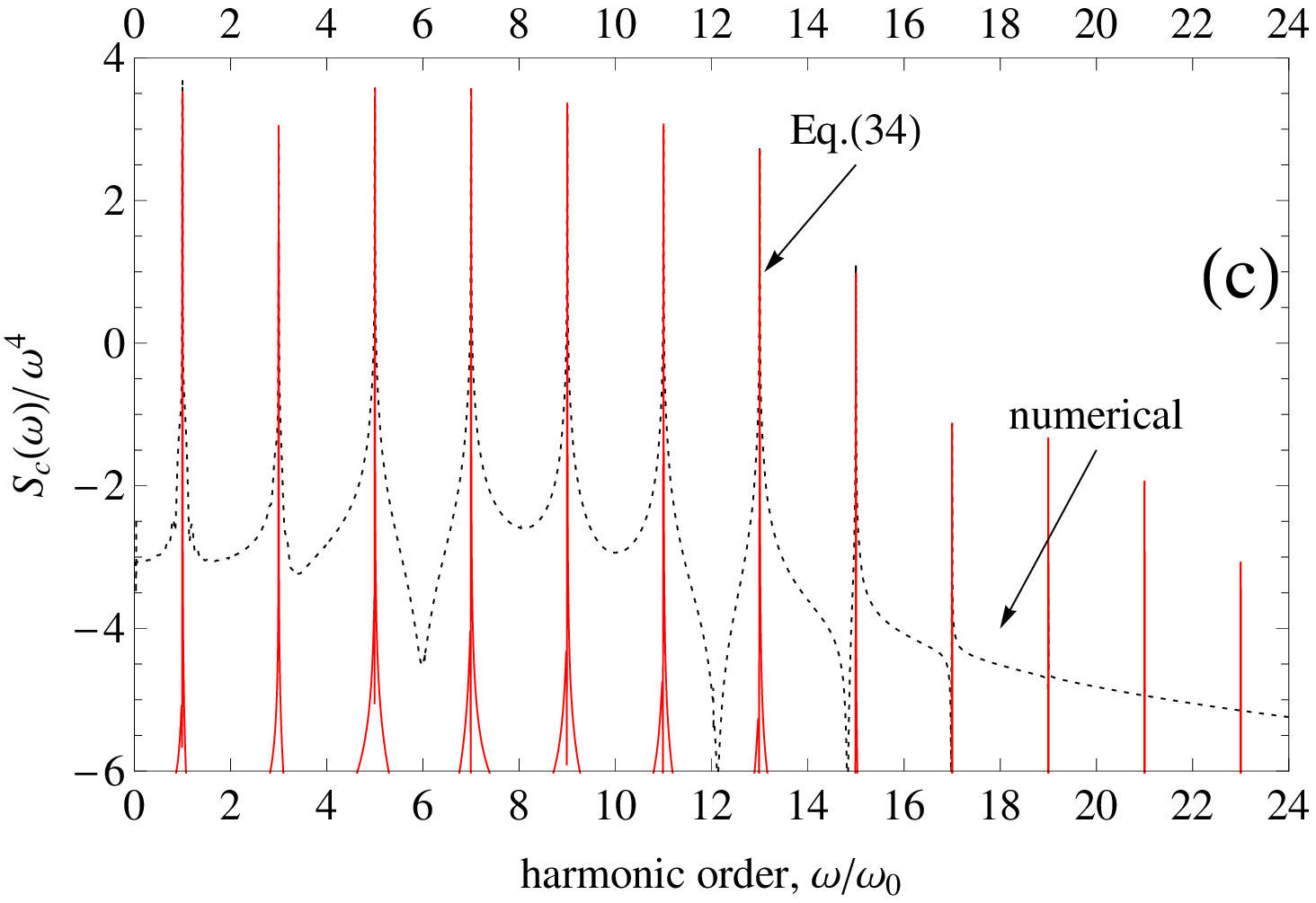}
\caption{(Color online) Nine-photon ($N=9$) resonance in the molecular ion $A_2^{4+}$ produced by the laser field of strength $E_0 = 0{.}099\text{ a.u.}$ and frequency $\omega_0 = 0{.}0754\text{ a.u.}$ ($\omega_0$ the same as in Fig.\ref{fig:2}). The description as in Fig.\ref{fig:2}.}\label{fig:3}
\end{figure}

As an example of a three-level system with degenerate upper states, we take the three lowest states in the hydrogen  atom: $\ket{1} = \ket{1S}$, $\ket{2} = \ket{2P}$ and $\ket{3} = \ket{2S}$. In this case, the excitation energy is $\omega_{21} = \omega_{31} = \frac{3}{8}\text{ a.u.}$, while the dipole matrix elements amount to $\mu_{12} = 0{.}745\text{ a.u.}$ and $\mu_{23} = -3\text{ a.u.}$, so $\mu_{12}/\mu_{23} = -0{.}248.$ For a square pulse, the system is expected to respond according to equations (\ref{eq:rwa}) and (\ref{eq:dip}) in which $\delta_N\rightarrow \bar{\delta}_N = \delta_N-(2\Delta_{f_N}+\Delta_{G_N})$, where $\Delta_{f_N}$ and $\Delta_{G_N}$ are to be calculated from equations (\ref{eq:ourdf}) and (\ref{eq:ourdG}), respectively. Thus, the N-photon resonance should occur at the light frequency $\omega_0 = \frac{1}{N}(\omega_{21}-(2\Delta_{f_N}+\Delta_{G_N}))$. We take a field allowed by the condition of applicability of equations (\ref{eq:rwa}) and (\ref{eq:dip}). For a given N-photon resonance, this condition reads $|a_N/\omega_0| = |N\frac{\mu_{12}}{\mu_{23}}J_N\left(\frac{M_R}{\omega_0}\right)|\ll 1$. Assuming $N=5$, for example, we find the ratio $|a_5/\omega_0|\leq 0{.}1$ if $M_R/\omega_0\leq 3{.}5$. So, we choose $M_R/\omega_0 = 1{.}5$ and obtain, for this field-strength parameter,  the following Stark shifts $\Delta_{f_5} = -0{.}00926\,\omega_0$ and $\Delta_{G_5} = -0{.}00646\,\omega_0$, giving the 5-photon resonant frequency $\omega_0 = 0{.}0754\text{ a.u.}$. The calculated $\omega_0$ leads to the electric field amplitude $E_0 = 1{.}5\omega_0/|\mu_{23}| = 0{.}0377\text{ a.u.}$, corresponding to the light intensity $1{.}42\cdot 10^{-3}\text{ a.u.}$ ($4{.}99\cdot 10^{13}\,\frac{\text{W}}{\text{cm}^2}$).For the above atom-field parameters, we show in Fig.2 the evolutions of the level populations and the coherent spectrum of the scattered light. Precisely, Fig.2a and Fig.2b show the evolution of the ground-state population, $|b_1(t)|^2$, and the evolution of the total population in the excited states, $|b_2(t)|^2+|b_3(t)|^2$. The bold solid curves in Fig.2a and Fig.2b were obtained from the simple analytic solution (\ref{eq:rwa}). The dotted curve with small fast oscillations in Fig.2a is the result of applying the improved approach of Avetissian \textit{et al.} when finding the fast parts $\beta_x$, $\beta_y$ and $\beta_z$ of $x_N$, $y_N$ and $z_N$, respectively (see the remark at the end of section \ref{sec:3}). In Fig.2b, the dotted curve with rapid oscillations comes from direct numerical integration of equations (\ref{eq:bamp}) for the field being turned on by 10 cycles according to $\sin^2(\omega_0 t/40)$, where $\omega_0 t/40\leq \pi/2$. As seen, the approximate analytic evolutions (Fig.2a) imitate quite well the numerical ones, though some difference in the periods of slow oscillations between the analytic and numerical curves must be noted. An important conclusion resulting from Fig.2a,b is that the simple solution (\ref{eq:rwa}) describes satisfactorily well the main trend observed in the numerical evolution, i.e., the possibility of achieving the total inversion of the population through multiphoton (5-photon) resonance. For the parameters of Fig.2, 225-cycle pulse ($0{.}45\text{ ps}$) turns out to be long enough to execute one trip in the population evolution, i.e., to move all the population from the ground state to the excited states and then back to the ground state. Under the condition of complete population inversion, Fig.2c shows the coherent part of the spectrum of scattered light (solid curve - Eq.(\ref{eq:dip}), dotted curve - numerical integration of Eq.(\ref{eq:bamp})). Except the height of the background, the analytic and numerical curves agree with respect to the number of peaks, their positions and relative heights.\\
\indent To present the results for a three-level system with nearly degenerate excited states we use the system parameters calculated by Gibson~\cite{gib2003} for his one-dimensional model of evenly charged homonuclear diatomic molecular ion $A_{2}^{4+}$, where $A=\text{N, O, I}$. It results from figures 10 and 11 of Gibson~\cite{gib2003} that, at the internuclear separation $3{.}5\text{ a.u.}$, the three lowest levels are so spaced that $\omega_{21} = 0{.}6685\text{ a.u.}$, $\omega_{32} = 0{.}0167\text{ a.u.}$ and, moreover, the dipole matrix elements are $\mu_{12} = 0{.}503\text{ a.u.}$ and $\mu_{23} = 3{.}033\text{ a.u.}$. Now, $\delta_N\rightarrow \widetilde{\delta}_N = \delta_N+\Delta-(2\Delta_{f_N}+\Delta_{G_N})$ in equations (\ref{eq:rwa}) and (\ref{eq:dip}), with $\Delta = \frac{\omega_{32}}{2}\left(1-J_0\left(\frac{2M_R}{\omega_0}\right)\right)$. As an illustration, let us consider the 9-photon resonance which is predicted at $\omega_0 = \frac{1}{9}(\omega_{21}+\Delta-(2\Delta_{f_9}+\Delta_{G_9}))$. We find that $|a_9/\omega_0|\leq 0{.}1$ provided that $M_R/\omega_0\leq 7{.}15$, so we choose $M_R/\omega_0=4$. For this value of $M_R/\omega_0$, we calculate $\Delta = 0{.}0069\text{ a.u.}$, $\Delta_{f_9} = -0{.}0155\,\omega_0$ and $\Delta_{G_9} = -0{.}014\,\omega_0$. These values recommend to take the laser frequency $\omega_0 = 0{.}0754\text{ a.u.}$, i.e., the same as that in the case of the 5-photon resonance in the hydrogen atom. The frequency $\omega_0$ gives the scaled separation $\omega_{32}/\omega_0 = 0{.}22$ and the peak electric field $E_0 = 0{.}099\text{ a.u.}$ translating into the light intensity $0{.}0098\text{ a.u.}$ ($3{.}44\cdot 10^{14}\,\frac{\text{W}}{\text{cm}^2}$). As found by Gibson \cite{gib2003}, in fields like that practically no other levels of a real molecular ion, besides the three discrete levels of the model, take part in the interaction. With the above ion-light parameters, we show in Fig.\ref{fig:3} the population evolutions and light scattering spectra. The qualitative conclusions resulting from Fig.\ref{fig:3} for the molecular ion match those drawn from Fig.\ref{fig:2} for the hydrogen atom. A quantitative difference between the figures is that, for the molecular ion, a longer pulse of 405 cycles ($0{.}82\text{ ps}$) is required to execute one complete trip in the population evolution (ground state$\rightarrow$excited state$\rightarrow$ground state). The other difference is a richer spectrum of scattered light including 12 components instead of 6 components met in the case of the hydrogen atom.\\
\begin{figure}[!ht]
\centering 
\includegraphics[width=0.9\columnwidth]{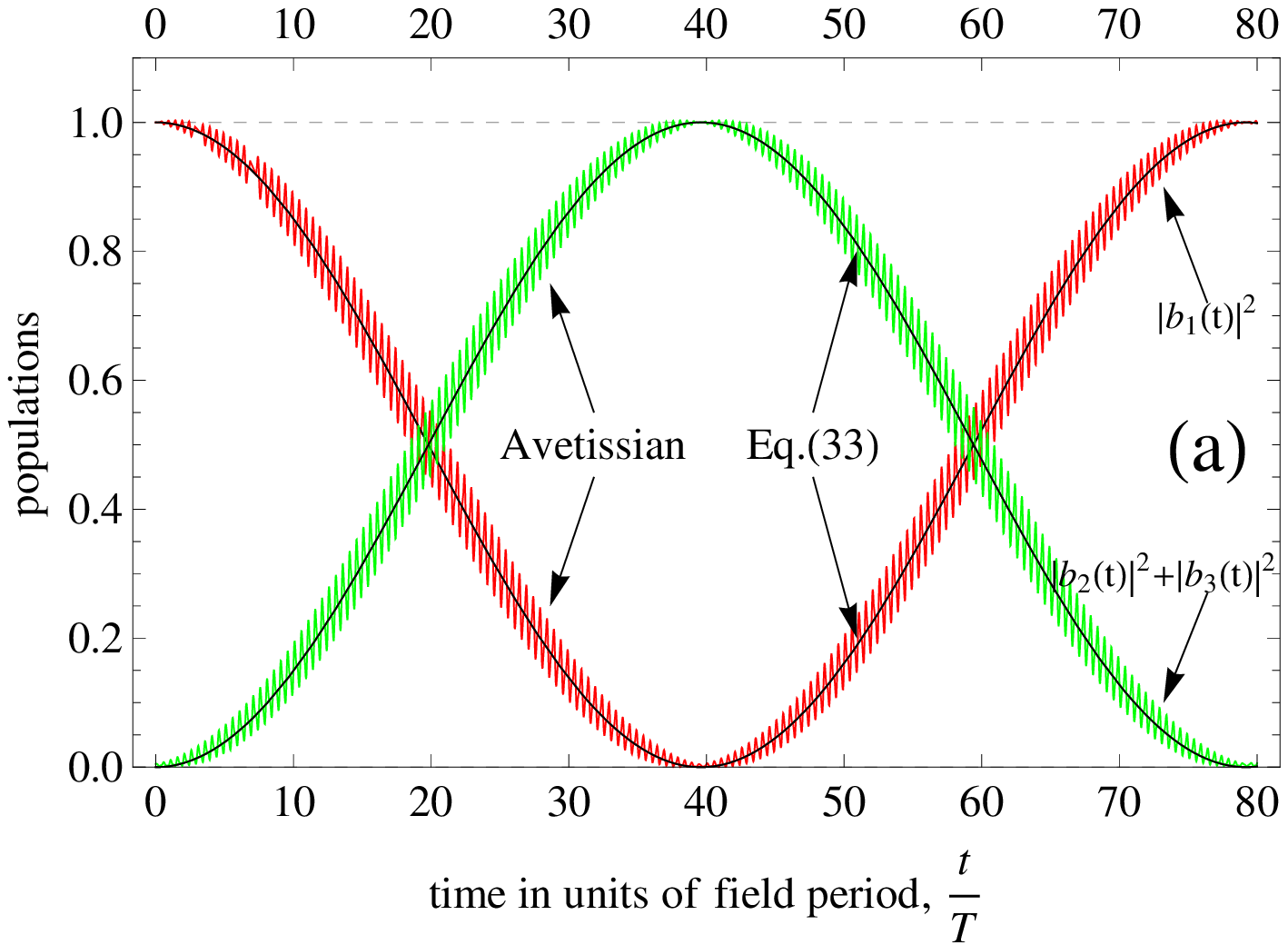}
\includegraphics[width=0.9\columnwidth]{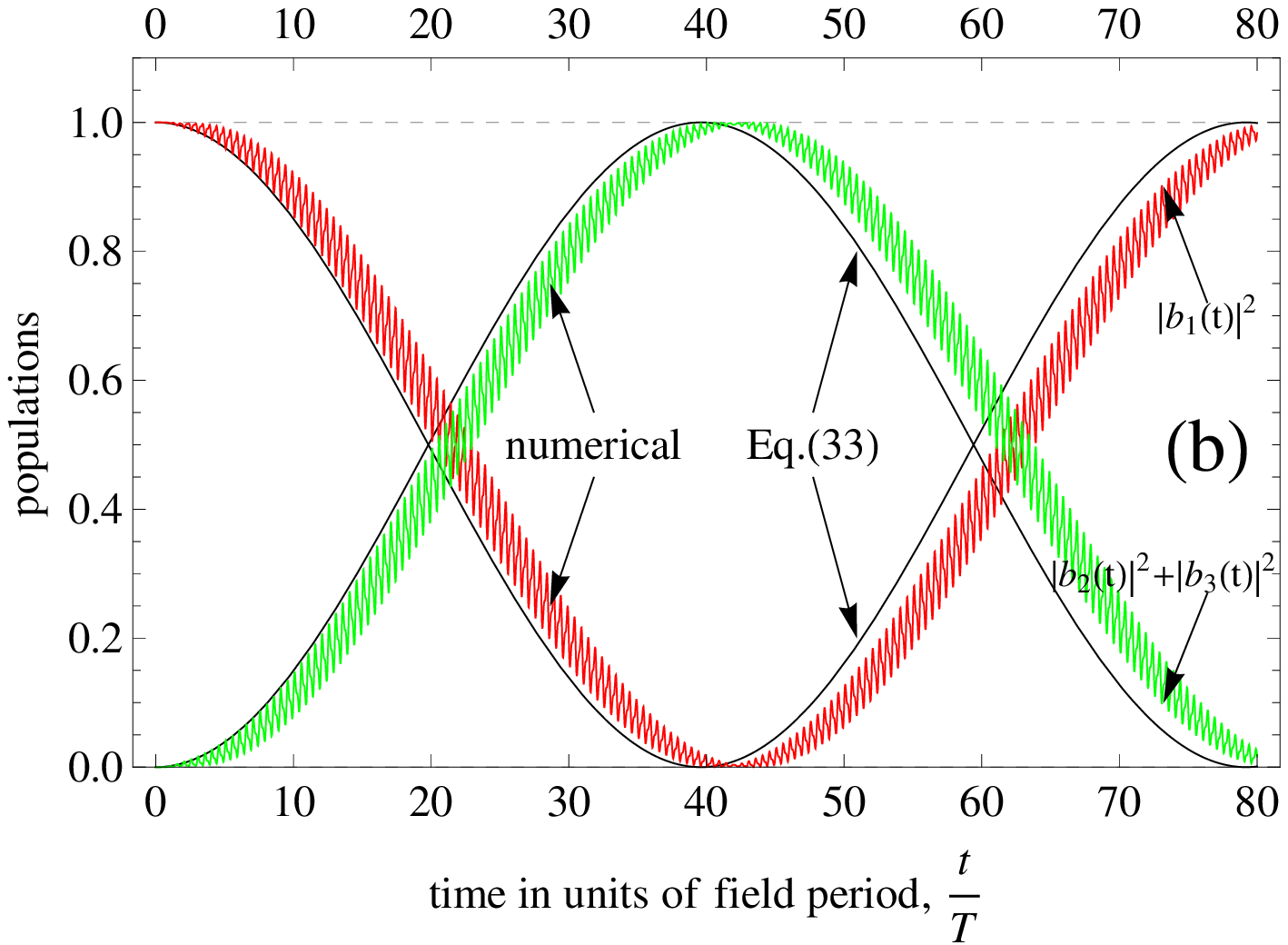}
\caption{(Color online) Population evolutions as in Fig.\ref{fig:2} but for three-photon ($N=3$) resonance in the hydrogen atom ($\omega_0 = 0{.}1255\text{ a.u.}$, $E_0 = 0{.}0314\text{ a.u.}$). In (b), the numerical curve is for field with 2-cycle turn-on time.}\label{fig:4}
\end{figure}
\indent We add Fig.\ref{fig:4} to show that, by lowering the order of resonance to $N=3$, specifically for the hydrogen atom, a better agreement was found between the analytically and numerically calculated population evolutions.\\
\indent We wish to stress that by ignoring the third states in the systems under study one obtains quite different results. For the hydrogen atom and the same field strength and frequency as in Fig.\ref{fig:2}, we found by numerical integration that the ground state in the two-level system is only weakly depopulated, and the spectrum includes only three components ($\omega_0$, $3\,\omega_0$, $5\,\omega_0$) with the 3-rd and 5-th harmonics being much weaker. Qualitatively the same result was found for the $A_2^{4+}$ ion and the same field strength and frequency as in Fig.\ref{fig:3}. The found weak depopulation of the ground ionic state resulted in a drastic reduction of the number of spectral components in the scattered light (to two only) and in lowering the height of the third harmonic. In Fig.\ref{fig:5} we show the numerically calculated spectra of scattered light for the above mentioned two-level systems obtained from our original three-level system (Fig.\ref{fig:sys}) by ignoring the third state of the same parity as that of the ground state \ket{1}.  Thus, the importance  of the third state is that the three-level system interacts much stronger with laser light giving stronger multiphoton excitation and harmonic generation, because of its strong coupling to the second state  by large dipole moment.

\begin{figure}[!ht]
\centering 
\includegraphics[width=0.9\columnwidth]{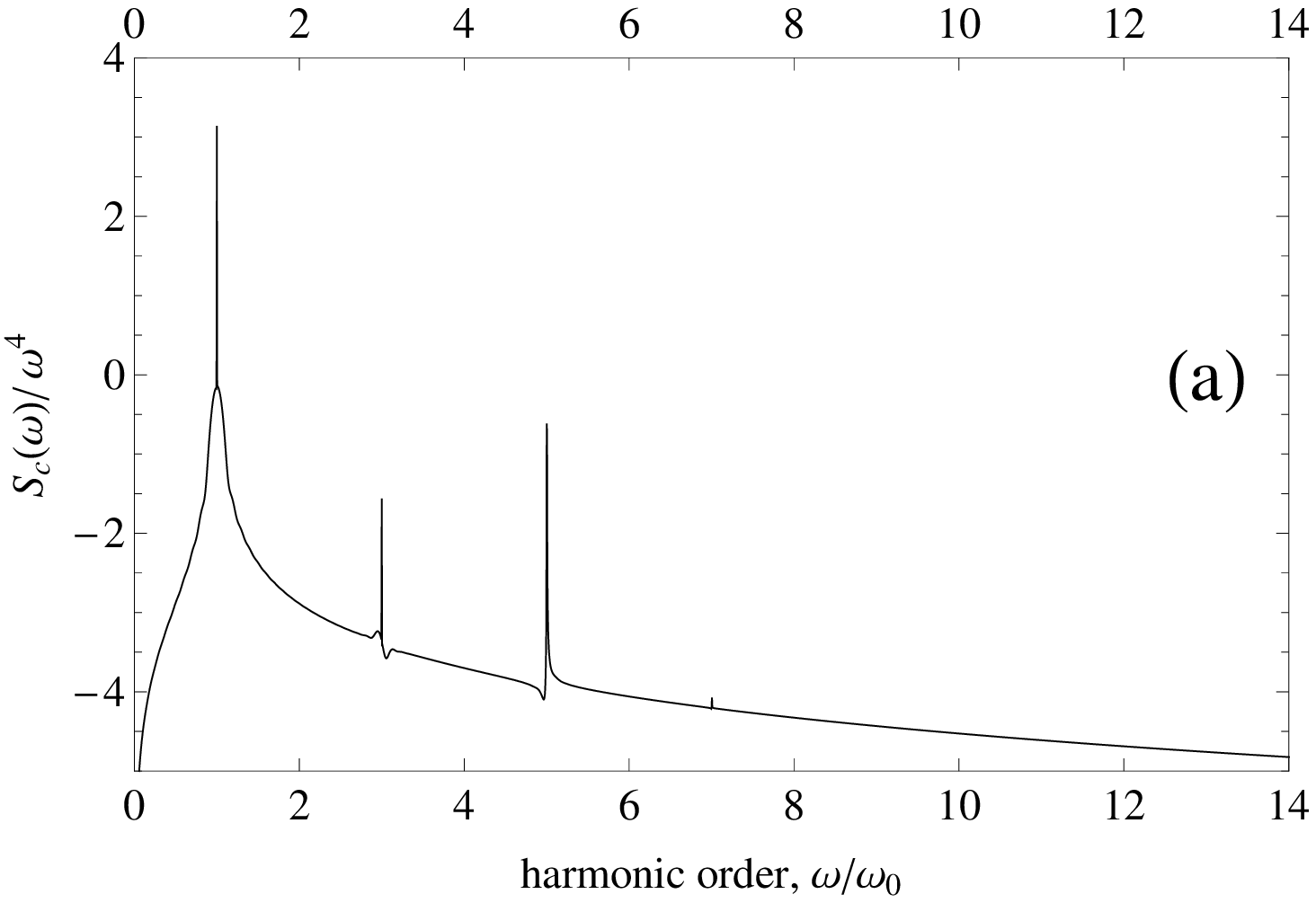}
\includegraphics[width=0.9\columnwidth]{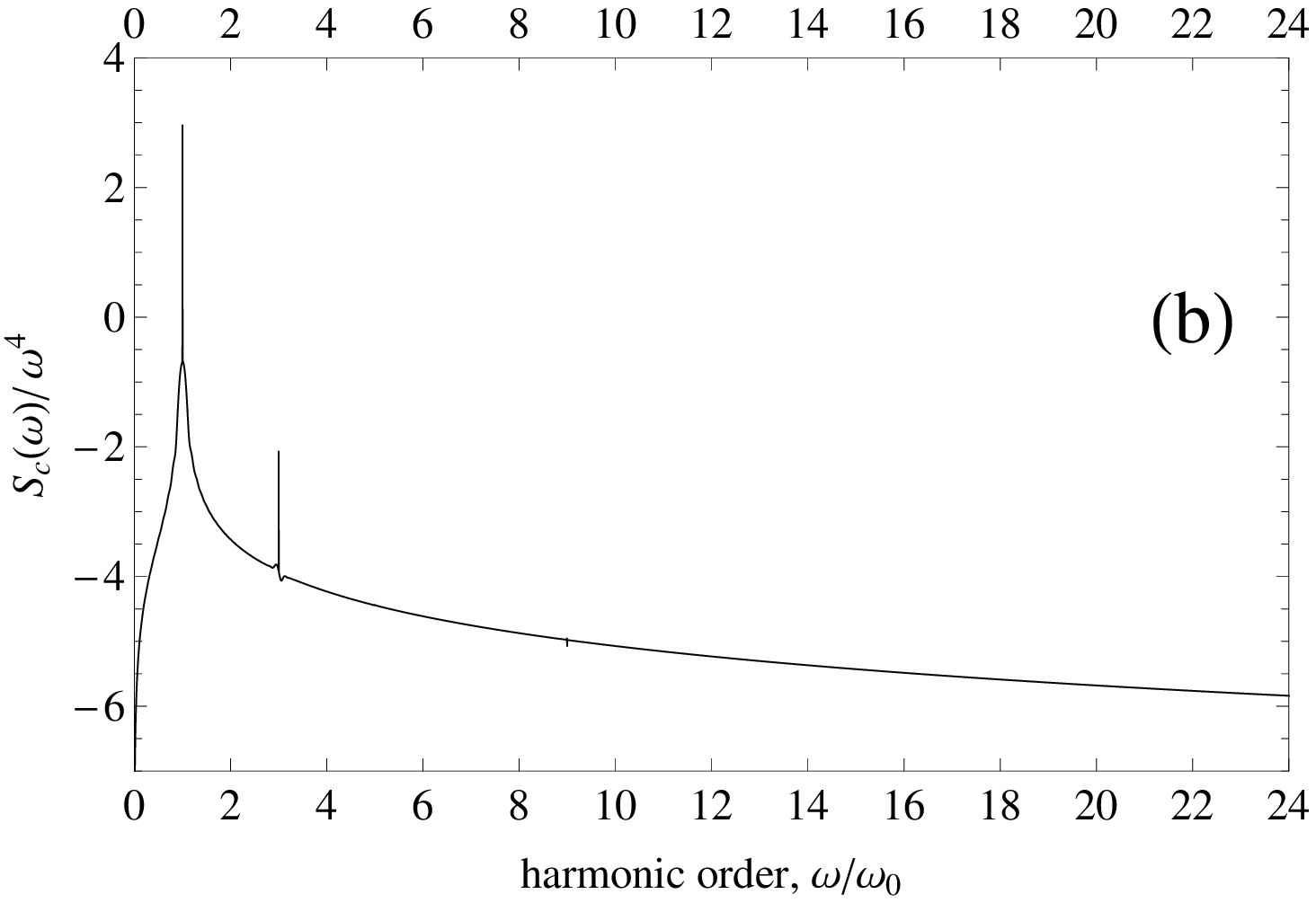}
\caption{(Color online) Numerical spectra of coherently scattered light obtained by neglecting the third state in the model shown in Fig.\ref{fig:sys}. (a) the spectrum from the two-level hydrogen atom and the  strength-frequency parameters as in Fig.\ref{fig:2}, (b) the spectrum from the two-level $A_2^{4+}$ ion and the strength-frequency parameters as in Fig.\ref{fig:3}.}\label{fig:5}
\end{figure}

\section{Summary}
In a specific three-level $\Gamma$-type system, we have treated analytically both multiphoton excitation and the accompanying generation of harmonics of the incident light. The specificity of the system was that it contained a pair of perfectly or nearly degenerate excited states separated by many photons from the ground state and, moreover, the dipole coupling between the excited states greatly exceeded that between the ground state and the lower excited state. Up to $100\%$ 5-photon excitation in the hydrogen atom and 9-photon excitation in the $A_{2}^{4+}$ molecular ion were found by subpicosecond laser pulses of moderate intensities ($5\cdot 10^{13}\,\frac{\text{W}}{\text{cm}^2}$ (hydrogen) and $3{.}44\cdot 10^{14}\,\frac{\text{W}}{\text{cm}^2}$ (molecular ion)) at the frequency $0{.}0754\text{ a.u.}$ ($2{.}05\text{ eV}$ photon energy). The calculated coherent spectrum of the scattered light was found to be composed of a number of odd-order harmonics and it was much richer and better pronounced than that from the two-level system obtained by rejecting one excited state of the same parity as that of the ground state. The analytic results agree, qualitatively at least, with those of numerical integration of the Schr\"odinger equation. The quantitative differences are that the numerically and analytically calculated populations oscillate slowly with slightly different frequencies, and the spectra of scattered light though coinciding in peaks have different heights of their backgrounds. A possible reason for different frequencies of slow oscillations is that, in the analytic calculations, we did not find sufficiently precisely the frequency $\omega_0$  generating multiphoton resonance. This supposition is maintained by the observation that changing slightly the calculated $\omega_0$ we were able to achieve better agreement between the analytic and numerical curves in figures \ref{fig:2} and \ref{fig:3}. Also, by diminishing the multiphoton order of resonance and, thus, the required laser intensity we found better agreement between the analytic and numerical results.

\end{document}